\documentclass[tradiabstract]{aa}

\newcommand{\bea}{\begin{eqnarray}}
\newcommand{\eea}{\end{eqnarray}}
\newcommand{\be}{\begin{equation}}
\newcommand{\ee}{\end{equation}}
\newcommand{\bc}{\begin{center}}
\newcommand{\ec}{\end{center}}
\newcommand{\ben}{\begin{enumerate}}
\newcommand{\een}{\end{enumerate}}
\newcommand{\bd}{\begin{description}}
\newcommand{\ed}{\end{description}}
\newcommand{\bmi}[1]{\begin{minipage}{#1 cm}}
\newcommand{\emi}{\end{minipage}}
\newcommand{\bmif}[1]{\begin{minipage}{#1\textwidth}}

\def\llabel#1{\label{sc:#1}  {#1}\hspace{0.5cm}}
\def\elabel#1{\label{eq:#1}\fbox{#1}}
\def\llabel#1{\label{sc:#1}}
\def\elabel#1{\label{eq:#1}}

\def\eck#1{\left\lbrack #1 \right\rbrack}

\def\rund#1{\left( #1 \right)}

\def\d{{\rm d}}

\def\Real{{\rm I\mathchoice{\kern-0.70mm}{\kern-0.70mm}{\kern-0.65mm}%
  {\kern-0.50mm}R}}
\def\C{\rm C\kern-.42em\vrule width.03em height.58em depth-.02em
       \kern.4em}

\def\bx#1{\leavevmode\thinspace\hbox{\vrule\vtop{\vbox{\hrule\kern1pt
        \hbox{\vphantom{\tt/}\thinspace{\bf#1}\thinspace}}
      \kern1pt\hrule}\vrule}\thinspace}

\def\vc#1{{\mbox{\boldmath$#1$\unboldmath}}}
{\catcode`\@=11
\gdef\SchlangeUnter#1#2{\lower2pt\vbox{\baselineskip 0pt \lineskip0pt
  \ialign{$\m@th#1\hfil##\hfil$\crcr#2\crcr\sim\crcr}}}
}

\def\ueber#1#2{{\setbox0=\hbox{$#1$}%
  \setbox1=\hbox to\wd0{\hss$\scriptscriptstyle #2$\hss}%
  \offinterlineskip
  \vbox{\box1\kern0.4mm\box0}}{}}

\def\bx#1{\leavevmode\thinspace\hbox{\vrule\vtop{\vbox{\hrule\kern1pt
        \hbox{\vphantom{\tt/}\thinspace{\bf#1}\thinspace}}
      \kern1pt\hrule}\vrule}\thinspace}

{\catcode`\@=11
\gdef\SchlangeUnter#1#2{\lower2pt\vbox{\baselineskip 0pt \lineskip0pt
  \ialign{$\m@th#1\hfil##\hfil$\crcr#2\crcr\sim\crcr}}}
}

\def\ts{\thinspace}

\usepackage{graphicx}
\usepackage{color}
\usepackage{txfonts}
\usepackage{natbib}
\usepackage{subfigure}
\usepackage{enumerate}
\usepackage{slashbox}
\begin{document}

   \title{Can one determine cosmological parameters from multi-plane
     strong lens systems?}

   \author{Peter Schneider \inst{1} 
          }

   \institute{Argelander-Institut f\"ur Astronomie, Universit\"at
     Bonn, Auf dem H\"ugel 71, D-53121 Bonn, Germany\\
    peter@astro.uni-bonn.de}


 
  \abstract
  { Strong gravitational lensing of sources with different redshifts
    has been used to determine cosmological distance ratios, which in
    turn depend on the expansion history. Hence, such systems are
    viewed as potential tools for constraining cosmological
    parameters. Here we show that in lens systems with two distinct
    source redshifts, of which the nearest one contributes to the
    light deflection toward the more distant one, there exists an
    invariance transformation that leaves all strong-lensing
    observables unchanged (except for the product of time delay and Hubble
    constant), generalizing the well-known mass-sheet transformation
    in single-plane lens systems. The transformation preserves the
    relative location of mass and light. All time
    delays (from sources on both planes) scale with the same factor --
    time-delay ratios are therefore invariant under the mass-sheet
    transformation. Changing 
    cosmological parameters, and thus distance ratios, is essentially
    equivalent to such a mass-sheet transformation. As an example, we
    discuss the double-source plane system SDSSJ0946+1006, which has
    recently been studied by Collett and Auger, and show that
    variations of cosmological parameters within reasonable ranges
    lead to only a weak mass-sheet transformation in both lens
    planes. Hence, the ability to extract cosmological information
    from such systems depends heavily on the ability to break the
    mass-sheet degeneracy.  }

  \titlerunning{Cosmology from multi-plane lensing?}

   \maketitle
%

\section{\llabel{Sc1}Introduction}
Strong gravitational lensing by galaxies is a powerful tool for
cosmological studies, in particular regarding precise mass estimates
of the inner region of galaxies, the angular structure of the mass
distribution (e.g., ellipticity and orientation), and mass
substructure \citep[see, e.g.,][and references
therein]{SaasFee3,Treu2010,Bart10}. In his pioneering paper, Refsdal
(\citeyear{Refsdal1964}) pointed out the possibility of determining
cosmological parameters from lensing, specifically the Hubble constant,
from measurements of time delays in lens systems. Whereas time delays
have been determined in some 20 lens systems by now, the accuracy of
the corresponding values of $H_0$ is difficult to judge, because of the
difficulty of reliably constraining the mass distribution of the lens: On
the one hand, the number of observational constraints can be
insufficient to constrain the mass distribution to sufficient
accuracy. On the other hand, \cite{FGS85} have found that there exists
a transformation of the mass distribution of the lens that leaves all
observables invariant, except for the product of time delay and
$H_0$. Thus, although very detailed studies of some lens systems have
been conducted \citep[see, e.g.,][]{Suyu2013a}, this mass-sheet
transformation (MST) poses a fundamental limitation of the accuracy of
the derived value for $H_0$ from gravitational lensing alone. Although
the degeneracy caused by the MST can be broken with additional
observations, such as stellar dynamics in the lens or some additional
information about the properties of the source (such as its
luminosity), the current accuracy on these quantities leads to
uncertainties of $H_0$ larger than those from other methods, and the
method may be biased. In practice, the degeneracy due to the MST is
broken by assumptions about the mass distribution, e.g., that it
follows a power law. A recent discussion of the impact of the MST on
$H_0$ determination can be found in \cite{SS13}.

A different approach to cosmology by strong-lens systems involves the
relative lensing strength for two sources at two different
redshifts. It is known that the classical MST is an exact
transformation of the lens mass distribution only for sources at a
single distance, and that it can be broken in principle when sources
at several redshifts are employed \citep{Bradac04}. However, it must
be stressed that this degeneracy-breaking assumes that there is no
second deflector along the line of sight to the sources.  In reality,
the lower-redshift sources have masses and consequently act as
additional lenses for higher-redshift sources.

The recent discovery of a strong galaxy-scale lens with two extended
multiple-image arc structures from two sources at vastly different
redshifts (SDSSJ0946+1006; Gavazzi et al.\ \citeyear{Gavaz08}) opened
up the possibility of studying a multi-plane lens system in great
detail. Collett \& Auger (\citeyear{CA14}; hereafter CA14) 
performed a detailed analysis of this system by
constructing a model that involves both the main lens in the
foreground of the two sources, and a smaller-mass deflector associated
with the lower-redshift source. The relative lens strength 
$\beta$ of the main lens on the two sources depends on distance
ratios, which in turn depend on the redshifts of lenses and sources
involved as well as on the distance-redshift relation in the
Universe. Since the latter is sensitive to the density parameters and
the equation of state parameter $w$ of dark energy, CA14 were
able to obtain constraints on $w$ from this lens system.

In this letter, we investigate whether this method can yield reliable
results. Specifically, we show in Sect.\ts\ref{sc:Sc2} that an analog
of the MST also exists for the case of two lens and two source planes,
as is the case for SDSSJ0946+1006. We then demonstrate in
Sect.\ts\ref{sc:Sc3} that a change of cosmological parameters that
leads to a change of the expected value of $\beta$ is equivalent to
such a generalized MST. We then study the amplitude of this equivalent
MST for a range of values of $w$, concluding that the transformation
amplitude across a plausible range of $w$ is indeed very small.  Since
the MST leads to a shape of the transformed density profile that is
different from the original, we conclude that this method heavily
relies on assumptions made for the shape of the mass profiles of the
lenses.

\section{\llabel{Sc2}Mass-sheet 
transformation for two lens planes}  
In this section we first summarize the lensing equations for the
multi-plane case (Sect.\ts\ref{sc:Sc2.1}), specify our notation, and
recall the MST in the single-plane case (Sect.\ts\ref{sc:Sc2.2})
before we derive the new MST for two lens and two source planes.

\subsection{\llabel{Sc2.1}Multi-plane lens equations}
We consider lenses and sources distributed along nearly the same
line of sight at $N$ different distances from us, characterized by
their redshifts $z_i$, or angular-diameter distances $D_i$ from us,
$1\le i\le N$ \citep[we largely follow the notation of][where more
details of 
the derivation are given]{SEF}. 
Perpendicular to the line of sight, we consider 
planes in which the sources at these distances are located, and onto
which the mass distribution of the deflecting masses are projected
(lens/source planes).  We denote by $D_{ij}$ the angular-diameter
distance of the $j$-th plane as seen from the $i$-th plane, with $1\le
i<j\le N$. The projected mass distribution in the $i$-th plane is
characterized by its surface mass density $\Sigma_i(\vc\xi)$ and gives
rise to a deflection angle ${\hat{\vc\alpha}}_i(\vc\xi_i)$, where
\be
{\hat{\vc\alpha}}_i(\vc\xi)={4 G\over c^2}
\int \d^2 \xi'\;\Sigma_i(\vc\xi')\,{\vc\xi -\vc\xi'\over |\vc\xi
  -\vc\xi'|^2} \;,
\elabel{hatalph-i}
\ee
and $\vc\xi_i=D_i \vc\theta_i$ is a transverse separation vector in the
$i$-th plane, with the corresponding unlensed angular position
$\vc\theta_i$.
The resulting propagation equations of a light ray follows solely from
geometry and the definition of angular-diameter distances,
\be
\vc\theta_j=\vc\theta-\sum_{i=1}^{j-1}{D_{ij}\over D_j} 
\hat{\vc\alpha}_i(D_i \vc\theta_i)=\vc\theta-\sum_{i=1}^{j-1}
\beta_{ij}\vc\alpha_i(\vc\theta_i)\;,
\elabel{LE}
\ee
where we set $\vc\theta\equiv\vc\theta_1$, scaled the deflection
angles $\hat{\vc\alpha}_i$ to the final source plane at $i=N$, i.e.,
\be
\vc\alpha_i(\vc\theta_i)={D_{iN}\over D_N}
\hat{\vc\alpha}_i(D_i \vc\theta_i)\;,\;\;\hbox{and defined}\;\;
\beta_{ij}={D_{ij}\over D_j}\,{D_{N}\over D_{iN}}
\elabel{betas}
\ee
as coefficients of relative distance ratios. Accordingly, we define
the dimensionless surface mass densities $\kappa_i(\vc\theta_i)=4\pi
G\,D_i\,D_{iN}\,\Sigma_i(D_i\vc\theta_i)/(c^2\,D_N)$, which satisfy
$\nabla \cdot \vc\alpha_i=2\kappa_i$.

\subsection{\llabel{Sc2.2}Summary of the single-plane MST} 
We briefly recall the MST for the single-lens plane, for which
$N=2$ \citep[see][and references therein]{FGS85,SS13}. Hence, we have
a single lens plane at redshift $z_1$ with deflection
$\vc\alpha(\vc\theta)\equiv \vc\alpha_1(\vc\theta_1)$ and
dimensionless surface mass density $\kappa(\vc\theta)\equiv
\kappa_1(\vc\theta_1)$, and a single source plane at redshift
$z_2$. 

If a mass distribution $\kappa(\vc\theta)$ can explain all the
observed lensing features, such as image positions, flux ratios,
relative image shapes, and time-delay ratios, of a source with
unlensed brightness profile $I^{\rm s}(\vc\theta_2)$, then the whole
family of mass distributions and corresponding scaled deflection angles
\be
\kappa_\lambda(\vc\theta)=\lambda\kappa(\vc\theta)+(1-\lambda)\; ;
\quad 
\vc\alpha_\lambda(\vc\theta)=\lambda
\vc\alpha(\vc\theta)+(1-\lambda)\vc\theta
\elabel{MST}
\ee
explains the lensed features equally well for a source of brighness
profile $I^{\rm s}_\lambda(\vc\theta_2)=I^{\rm s}(\vc\theta_2/\lambda ) $.
Hence, the transformed brightness distribution in the source plane is
rescaled by a factor $\lambda$. This rescaling of the source plane
implies that the magnification of images is changed,
$\mu_\lambda=\mu/\lambda^2$, but that can only be observed if the
luminosity or physical size of the source is known (standard candle
or standard rod). In most cases, this information is unavailable, so
that the magnification cannot be obtained from observations, whereas
magnification (and thus flux) ratios are invariant under the MST. The
time delay between pairs of images is changed under the MST, so that a
time delay measurement can be employed to break the degeneracy caused
by the MST, provided the Hubble constant (and the density parameters
of our Universe) are assumed to be known. 

\subsection{\llabel{Sc2.3}MST for two source and lens planes} 
We now study whether a similar invariance transformation exists
for sources at two different redshifts $z_2$ and
$z_3$, with lenses at redshift $z_1$ and $z_2$. In particular, we
consider that the closer source at $z_2$ is associated with mass that
deflects light rays from the more distant source. By specializing the
equations of Sect.\ts\ref{sc:Sc2.1} to the case $N=3$, we obtain the
pair of relations
\be
\vc\theta_2=\vc\theta-\beta\,\vc\alpha_1(\vc\theta)\;; \;\; 
\vc\theta_3=\vc\theta-\vc\alpha_1(\vc\theta)-\vc\alpha_2(\vc\theta_2)\;,
\elabel{LE3}
\ee
where we defined $\beta\equiv \beta_{12}$ (see
Eq.\ts\ref{eq:betas}). A mass-sheet transformation is defined as a
change of the deflection angles (and correspondingly the lensing mass
distributions) such that the lens equations remain invariant, with only
a uniform isotropic scaling in the source planes. Considering the
first of Eq.\ts(\ref{eq:LE3}), a MST for the first source plane is obtained
by applying the single-lens plane MST to the effective deflection
$\beta\vc\alpha_1$, i.e., by setting (here and in the
  following, a prime denotes transformed quantities)
\be
\vc\alpha_1'(\vc\theta)=\lambda \vc\alpha_1(\vc\theta)
+{1-\lambda\over\beta}\vc\theta\;,
\elabel{al1}
\ee
after which the transformed position at $z_2$ becomes
\be
\vc\theta_2'=\lambda\vc\theta-\lambda\beta\vc\alpha_1(\vc\theta)
=\lambda\vc\theta_2\;,
\ee
i.e., the required uniform isotropic scaling. The question now is
whether we can find a transformation of the deflection angle
$\vc\alpha_2$ such that the second of Eq.\ts(\ref{eq:LE3}) also remains
invariant up to a scaling of $\vc\theta_3$. If
$\vc\alpha_2'(\vc\theta_2')$ denotes the transformed deflection angle,
then the transformed lens equation reads
\be
\vc\theta_3'=\vc\theta-\lambda\vc\alpha_1(\vc\theta)
-{1-\lambda\over\beta}\vc\theta-\vc\alpha_2'(\lambda\vc\theta_2)\;.
\ee
Requiring that $\vc\theta_3'=\nu_3 \vc\theta_3$, corresponding to
uniform scaling with the factor $\nu_3$, we obtain
\be
{\beta+\lambda-1\over \beta}\vc\theta-\lambda\vc\alpha_1(\vc\theta)
-\vc\alpha_2'(\lambda\vc\theta_2)
=\nu_3\eck{\vc\theta-\vc\alpha_1(\vc\theta)-\vc\alpha_2(\vc\theta_2)}\;.
\elabel{con1}
\ee
To satisfy this equation, the
term $\vc\alpha_2'$ on the l.h.s. of Eq.\ts(\ref{eq:con1}) must
contain the term $\nu_3\vc\alpha_2(\vc\theta_2)$. Therefore, we set
\be
\vc\alpha_2'(\vc\theta_2')=\nu_3 \vc\alpha_2(\vc\theta_2'/\lambda)
+K_2\vc\theta_2' = 
\nu_3 \vc\alpha_2(\vc\theta_2'/\lambda)+K_2\lambda\eck{\vc\theta
-\beta\vc\alpha_1(\vc\theta)} \;,
\elabel{con2}
\ee
where in the second step we used $\vc\theta_2'=\lambda\vc\theta_2$ and
the first of Eq.\ts(\ref{eq:LE3}). Here, $K_2$ is a constant, to be
constrained later. Inserting Eq.\ts(\ref{eq:con2}) into
Eq.\ts(\ref{eq:con1}), we 
see that the terms involving $\vc\alpha_2$ vanish. Comparing the
remaining terms proportional to $\vc\alpha_1$ and $\vc\theta$, we find
the pair of constraints
$\lambda-K_2 \lambda\beta=\nu_3$ and
$\beta+\lambda-1-K_2 \lambda\beta=\beta \nu_3$,
which have the unique solution
\be
\nu_3=1\; ; \;\; K_2={\lambda-1 \over \lambda \beta}\;
\Rightarrow\;\;
\vc\alpha_2'(\vc\theta_2')=\vc\alpha_2(\vc\theta_2'/\lambda)  
+{\lambda-1 \over \lambda \beta}\vc\theta_2' \;.
\elabel{con4}
\ee
Hence we obtain a solution of Eq.\ts(\ref{eq:con1}) and accordingly a
transformation of the mass distributions in both lens planes, which
leads to at most a uniform isotropic scaling of the source planes. 

The transformation of the first lens plane is a normal MST, in that
the deflection angle (and thus the surface mass density) is scaled by
an overall factor $\lambda$, and a uniform density
$K_1=(1-\lambda)/\beta$ is added, leading to a scaling of the first
source plane by a factor $\lambda$. This scaling equally
  applies to the light distribution and the mass distribution, as seen
  by Eq.\ts(\ref{eq:con2}). Thus, after the MST, both the light and the
mass in the plane $i=2$ are transformed in exactly the same way. In
particular, this means that if the original model has a mass component
centered on a light component, the same remains true after the MST,
but both are located at a position that differs by a factor of
$\lambda$.

Condition (\ref{eq:con2}) furthermore implies that the transformed
deflection in the second lens plane is a scaled version of the
original deflection, plus a contribution from a uniform mass sheet $K_2$. 
The implication of the scaling of the first term in Eq.\ts(\ref{eq:con2})
on the corresponding mass distribution can be seen as follows:
If  
\be
\vc\alpha(\vc\theta|\kappa(\vc\theta))  
={1\over \pi}\int\d^2 \theta'\; \kappa(\vc\theta')\,
{\vc\theta-\vc\theta'\over |\vc\theta-\vc\theta'|^2}
\ee
is the deflection caused by the mass distribution $\kappa$, then
\be
\vc\alpha(\vc\theta/\lambda|\kappa(\vc\theta))
= \vc\alpha(\vc\theta|\lambda^{-1}\kappa(\vc\theta/\lambda)) \;.
\ee
Hence, the transformed mass distribution $\kappa_2'$ is a scaled
version of the original one, multiplied by a factor $\lambda^{-1}$,
plus a uniform mass sheet. It must be stressed here that the phrase
`adding a uniform mass sheet' corresponds to a global interpretation
of the MST; however, only a relatively small inner region of the lens
is probed by strong lensing, and hence the MST needs to apply only
locally. Its main effect is the change of the local slope of the mass
profile near the Einstein radius of a lens, making it flatter
(steeper) for $\lambda<1$ ($\lambda>1$). The global interpretation of
the mass sheet as an `external convergence', relating it to the
large-scale environment of the lens which can be probed by the
observed density field of galaxies \citep[e.g.,][]{Wong11, Coll13,
  Greene13}, constitutes an extrapolation over a vast range of scales.

Surprisingly, the second source plane remains unscaled under this MST,
because of $\nu_3=1$. Hence, the MST implies no change in the mapping of the
second source plane, including no change in the magnification matrix. 
Whereas simple algebra has straightforwardly led to this result, the
geometrical reason for this appears unclear. 

If the MST is such that the scaling in the first lens plane
corresponds to an additional focusing, which means $\lambda<1$, so
that the uniform mass sheet has positive convergence, then the mass
sheet in the second lens plane has negative convergence, since the
sign of $K_2$ is opposite to that of $1-\lambda$. In particular,
$\lambda =1$ implies $K_2=0$, so there is no MST that leaves the
first lens plane invariant and only affects the second one. It must be
stressed that the MST leaves the global lens mapping invariant
and thus applies to sources of (in principle) arbitrary extent. This
is quite different from other modifications of the lens mass
distribution \citep[e.g.,][and references
therein]{Coe08,Liesenborgs08,Liesenborgs2012}, which apply to a 
discrete set of isolated small images of sources.

\subsection{\llabel{Sc2.4}Time delay}
We now consider the impact of the MST on the time delay. For sources
at $z_2$, the MST is a normal single-plane MST, and the time delays
are changed by a factor $\lambda$. For sources at $z_3$, we use the
expression for the light travel-time from $\vc\theta_3$ via
$\vc\theta_2$ and $\vc\theta$ to the observer, as given in \cite{SEF},
\be
T(\vc\theta,\vc\theta_2,\vc\theta_3)=
\sum_{i=1}^2 {1+z_i\over c}\,
{D_i D_{i+1}\over D_{i,i+1}}\,
\tau_{i,i+1}(\vc\theta_i,\vc\theta_{i+1})\; ,
\ee
where $\tau_{i,i+1}=(\vc\theta_i-\vc\theta_{i+1})^2/2
  -\beta_{i,i+1}\psi_i(\vc\theta_i)$
is the Fermat potential corresponding to neighboring planes, with
$\beta_{12}\equiv\beta$ and $\beta_{23}=1$, and
$\psi_i(\vc\theta_i)$ is the deflection potential with
$\nabla\psi_i=\vc\alpha_i$. We now consider how $T$ behaves under an
MST. The scalings (\ref{eq:al1}) and (\ref{eq:con2}) of $\vc\alpha_i$ imply
that 
\be
\psi_1'(\vc\theta)=\lambda \psi_1(\vc\theta)+{1-\lambda\over
  2\beta}\vc\theta^2 \; ;
\;\;
\psi_2'(\vc\theta_2')=\lambda \psi_2(\vc\theta_2'/\lambda)+{K_2\over 2}
{\vc\theta_2'}^2 \;.
\ee
Together with $\vc\theta_2'=\lambda\vc\theta_2$ and
$\vc\theta_3'=\vc\theta_3$, we obtain
\bea
\tau_{12}'\!\!\!\!&=&\!\!\!\!{(\vc\theta-\lambda\vc\theta_2)^2\over 2}
-\beta\rund{\lambda\psi_1(\vc\theta)+{1-\lambda\over
    2\beta}\vc\theta^2} \nonumber \\
\!\!\!\!&=&\!\!\!\!\lambda\eck{ {(\vc\theta-\vc\theta_2)^2\over
    2}-\beta\psi_1(\vc\theta)}
+{\lambda (\lambda-1)\over 2}\vc\theta_2^2 \;;\nonumber \\
\tau_{23}'\!\!\!\!&=&\!\!\!\!{(\lambda\vc\theta_2-\vc\theta_3)^2\over 2}
-\lambda\psi_2(\vc\theta_2)-{K_2\over 2}\lambda^2\vc\theta_2^2 \\
\!\!\!\!&=&\!\!\!\! \lambda \eck{ {(\vc\theta_2-\vc\theta_3)^2\over
    2}-\psi_2(\vc\theta_2)}
+{\lambda^2-\lambda\over 2}\rund{1-{1\over \beta}}\vc\theta_2^2
+{1-\lambda\over 2}\vc\theta_3^2 \;.\nonumber
\eea
Hence, the light travel-time function transforms as
\bea
T'(\vc\theta,\vc\theta_2',\vc\theta_3')
\!\!\!\!&=&\!\!\!\!
\lambda T(\vc\theta,\vc\theta_2,\vc\theta_3)
+{1+z_2\over c}{D_2 D_3\over D_{23}}{1-\lambda\over 2}\vc\theta_3^2\nonumber \\
+ {\lambda(\lambda-1)\over 2 c}\vc\theta_2^2
\!\!\!\!\!\!\!\!\!\!\!\!&&\!\!\!\eck{(1+z_1)
{D_1 D_2\over D_{12}}+(1+z_2)
{D_2 D_3\over D_{23}}\rund{1-{1\over \beta}}}\;.
\elabel{Ttrans}
\eea
The second term in Eq.\ts(\ref{eq:Ttrans}) only depends on the source
position $\vc\theta_3$; therefore, this term does not contribute to
the time delay, which is obtained as the difference of $T$ between
images.  The third term of Eq.\ts(\ref{eq:Ttrans}) vanishes, since the
expression in the bracket is zero because of relations between the $D$'s
-- see \cite{SEF}. We thus find that all time delays in a
two-source plane lens scale with $\lambda$ under an MST, for sources
on 
both source planes. Hence, time-delay ratios do not break the
degeneracy of the MST; on the other hand, any time delay from a source
in either source plane breaks it, provided $H_0$ is
assumed to be known.

\section{\llabel{Sc3}Cosmology from two-source 
plane lensing?}  
In the light of the new MST, we now consider the possibility of
constraining cosmological parameters from 
two-source plane lenses. In the approach of CA14, the
sensitivity to cosmology is due to the distance ratio parameter
$\beta$, which depends on the density parameters and the dark energy
e.o.s. parameter $w$. Different cosmological models yield different
distance-redshift relations, and thus different $\beta$. Thus let
$\beta$ correspond to a fiducial cosmological model, and $\beta'$ to a
model with different parameters. We examine below whether we can find
deflection angles $\vc\alpha'_i(\vc\theta'_i)$ such that the lens
mappings are unchanged between the fiducial and modified models, up to
a uniform scaling (by a factor $\nu_i$) in the lens/source
planes. Thus we require for the first lens equation
\be
\vc\theta_2'=\vc\theta-\beta'\vc\alpha_1'(\vc\theta)
=\nu_2\vc\theta_2=\nu_2\eck{\vc\theta-\beta\vc\alpha_1(\vc\theta)}
\;. 
\elabel{C-LE1}
\ee
With the ansatz $\vc\alpha_1'(\vc\theta) =
\lambda\vc\alpha_1(\vc\theta)+K_1\vc\theta$, we obtain by comparing
terms in Eq.\ts(\ref{eq:C-LE1}) proportional to $\vc\alpha_1$ and
$\vc\theta$ the two relations $\beta'\lambda=\nu_2\beta$ and
$1-\beta' K_1=\nu_2$, with solutions $\nu_2=\beta'\lambda/\beta$ and
$K_1=1/\beta'-\lambda/\beta$. The requirement for the second lens
equation reads
\be
\vc\theta_3'=\vc\theta-\vc\alpha_1'(\vc\theta)-\vc\alpha_2'(\vc\theta_2')
=\nu_3\vc\theta_3=\nu_3\eck{\vc\theta-\vc\alpha_1(\vc\theta)
-\vc\alpha_2(\vc\theta_2)} \;.
\elabel{C-LE2}
\ee
Inserting the ansatz
\[
\vc\alpha_2'(\vc\theta_2')=\nu_3\vc\alpha_2(\vc\theta_2'/\nu_2)
+K_2 \vc\theta_2'
=\nu_3\vc\alpha_2(\vc\theta_2'/\nu_2)+K_2 \nu_2
\eck{\vc\theta-\beta\vc\alpha_1(\vc\theta)}
\elabel{C-LE3}
\]
into Eq.\ts(\ref{eq:C-LE2}), the comparison of terms propotional to
$\vc\alpha_1$ and $\vc\theta$ yields the equations
$\lambda-K_2\nu_2\beta=\nu_3$ and $1-K_1-K_2\nu_2=\nu_3$, which
have the solutions
\be
\nu_3={\beta\over\beta'}\,{1-\beta'\over 1-\beta}\; ;
\;\;
K_2={1\over \beta'}\rund{1-{\beta\over\lambda\beta'}\,{1-\beta'\over
    1-\beta}} \;.
\ee
Hence we see that a change of cosmology changes the lens equations in
a similar way as an MST. Even with a change in $\beta$, we retain the
freedom of a one-parameter family of mass models that leave the lens
mapping invariant, up to a uniform scaling. In  contrast to the MST
discussed in Sect.\ts\ref{sc:Sc2.3}, here a non-trivial scaling of
the second lens plane is implied, where $\nu_3$ solely depends on
change of $\beta$.

Three special choices of $\lambda$ are worth to be discussed separately:
{\it (1) No mass sheet in the first lens plane (NMS1):} For
$\lambda=\beta/\beta'$, $K_1=0$ and $\nu_2=1$, so that there is no
mass sheet in the first lens and no scaling in the first source/second
lens plane is implied.  The mass sheet in the second lens plane then
becomes $K_2=(\beta'-\beta)/[\beta'(1-\beta)]$. For four different
variations around a fiducial cosmological model, all with zero spatial
curvature, we give the values of $\beta'$, $K_2$ and $\nu_3$ in
Table\ts\ref{tab1}, using the redshifts for the system SDSSJ0946+1006
(setting the uncertain redshift of the second source to be
$z_3=2.4$). {\it (2) No mass sheet in the second lens plane (NMS2):}
Setting $\lambda=\nu_3$, we derive $K_2=0$, $\nu_2=(1-\beta')/(1-\beta)$,
and $K_1= (\beta'-\beta)/[\beta'(1-\beta)]$, so that $\lambda+K_1=1$.
Hence, in this case, there is no mass sheet in the second lens plane,
and the one in the first lens plane is the same as $K_2$ was for
NMS1.  {\it (3) Equal-mass scaling (EMS):} Choosing sheets of equal
density in both lens planes, we derive
$\lambda=(\beta/\beta')\sqrt{(1-\beta')/(1-\beta)}$,
$K_1=K_2=(1-\sqrt{(1-\beta')/(1-\beta)})/\beta'$, and $\nu_2=
\sqrt{(1-\beta')/(1-\beta)}$. For these choices of $\lambda$, we
also list some quantities for the different cosmological parameters in
Table\ts\ref{tab1}.

\begin{table}
\caption{MST parameters for modified cosmological parameters.}

\label{tab1}
\bc
\begin{tabular}{|c|rrrr|}
\hline
$\Omega_{\rm m}$ & 0.3 & 0.3 & 0.2 & 0.4 \\
$w$ & $-0.5$ & $-1.5$ & $-1$ & $-1$ \\
\hline
\hline
$\beta'$ & 0.697 & 0.720 & 0.710 & 0.706 \\
$\nu_3$ & 1.058 & 0.944 & 0.993 & 1.010 \\
\hline
$\lambda$ & 0.983 & 1.017 & 1.002 & 0.997 \\
$K_2 $ & $-0.058$ &0.056 & 0.007 & $-0.010$ \\
\hline
$\lambda$ & 1.058 & 0.944 & 0.993 & 1.010 \\
$K_1 $ & $-0.058$ &0.056 & 0.007 & $-0.010$ \\
$\nu_2$   & 1.040 & 0.960 & 0.995 & 1.007 \\
\hline
$\lambda$ & 1.037 & 0.963 & 0.996 & 1.006 \\
$K_1=K_2$ & $-0.029$ & 0.028 & 0.003 & $-0.005$ \\
$\nu_2$ & 1.020 & 0.980 & 0.998 & 1.003\\
\hline
\end{tabular}
\ec
Notes: For the redshifts of the two lens plane system
  SDSSJ0946+1006, i.e., $z_1=0.222$, $z_2=0.609$, $z_3=2.4$, we
  consider four different variations of the fiducial flat cosmological
  model (first block), for which
  $\Omega_{\rm m}=0.3$, $w=-1$, and thus $\beta=0.708$. The second
  block lists $\beta'$ and $\nu_3$ for these cosmologies. Then three
  special choices of $\lambda$ are considered, as described in the
  text, and listed in blocks 3 (NMS1), 4 (NMS2), and 5 (EMS).

\end{table}

From the table, we see that variations of the cosmological parameters
within generous plausible ranges yield only small deviations of
$\beta'$ from $\beta$. Correspondingly, the required MSTs are small;
for the three special choices discussed above they imply mass sheets
with $|K_i|<0.06$. In particular, for EMS, the mass-sheet densities are
smaller than $0.03$. Such low values of $K_i$ are perfectly acceptable,
even if fairly accurate measurements of the velocity dispersion of the
lenses were available \citep[see][for a more detailed
discussion]{SS13}. In particular, if the original mass distribution
were a 
power law, the transformed ones $\kappa_i'$ would deviate only very
little from a power law over the range where multiple images are
formed. 

\section{Discussion}
We have shown that a lens system with two 
lens and two source planes admits a
mass-sheet transformation that leaves the lens mapping invariant up
to a uniform scaling in the source and lens plane(s). Furthermore, we
demonstrated in the previous section that a change of cosmological
parameters is essentially equivalent to an MST. A mass
  sheet acts like a magnifying glass, changing apparent distances,
  much in the same way as a different expansion history changes the
  distance-redshift relation.
For the particular case
of SDSSJ0946+1006 \citep{Gavaz08}, we have shown that the changes of
the mass distribution in the second lens plane are very small when
the cosmological parameters are changed within currently acceptable
ranges.

As explicitly stated in CA14, the cosmological constraints they obtained
are based on the assumption of a power-law mass distribution in either
lens planes. This assumption about the functional
  form of the density profile formally breaks the degeneracy
  implied by the MST, but there are no good reasons to assume that the
  profiles of galaxies are {\it exactly} described by a power law
  \citep[see][]{SS13}.  Indeed, the case NMS1 discussed above implies
no added mass sheet in the first lens plane, thus preserving the
power-law property, whereas the case EMS corresponds to only small
$K_i$, yielding only marginal modifications of the power-law mass
profile.  The MST also for multiple
source and lens planes provides a degeneracy in the determination of
lens-mass distributions and cosmological distance ratios in lens
systems, 
and needs to be accounted for in future studies.

\begin{acknowledgements}
The author thanks Dominique Sluse, Tom Collett, Sherry Suyu, and an
anonymous referee for helpful
discussions and comments on the manuscript. This work was supported
in part by the Deutsche Forschungsgemeinschaft under the TR33 `The
Dark Universe'. 
\end{acknowledgements} 

\bibliographystyle{aa}
\bibliography{MSDbib}



\end{document}